\documentclass[11]{article}
\usepackage{feynmp}
\usepackage{hyperref}	
\usepackage{amsmath}	
\usepackage{graphicx}	
\usepackage{fullpage}	
\graphicspath{{picture/}}
\usepackage{ amssymb }
\usepackage{caption}
\begin{document}


\title{Surface Delta Interaction in the $g_{7/2}-d_{5/2}$ Model Space }

\author{Xiaofei Yu and Larry Zamick\\
 \textit{$^{1}$Department of Physics and Astronomy, Rutgers University,
Piscataway, New Jersey 08854, USA} }

\date{\today}
\maketitle

\begin{abstract}
Using an attractive surface delta interaction we obtain wave functions
for 2 neutrons (or neutron holes) in the $g_{7/2}-d_{5/2}$ model
space. If we take the single particle energies to be degenerate we
find that the $g$ factors for $I=2,4$ and $6$ are all the same $G(J)=g_{l}$,
the orbital $g$ factor of the nucleon. For a free neutron $g_{l} =0$
so in this case all $2$ particles or $2$ holes' $g$ factors are equal to zero.
Only the orbital part of the $g$-factors contribute - the spin
part cancels out. We then consider the effects of introducing a single
energy splitting between the $2$ orbits. We make a linear approximation
for all other $n$ values.
\end{abstract}

\section*{ Surface Delta Interaction}

The surface delta interaction (SDI) of Green and Moszkowski \cite{Green}
and Arvieu and Moszkowski \cite{Arvieu}has proven to be a very useful schematic
interaction. It can be used to find the hidden simplicity in complex
calculations. This interaction has been extensively discussed in Talmi's
book \cite{Talmi}.

The matrix element of the SDI interaction can be written as follows:
\begin{equation}
<[j_{1}j_{2}]SDI[j_{3}j_{4}>=C_{0}f(j_{1},j_{2})f(j_{3},j_{4})
\end{equation}
 where we have 
\begin{equation}
f(j_{1},j_{2})=(-1)^{j_{2}+\frac{1}{2}}\sqrt{\frac{(2j_{1}+1)(2j_{2}+1)}{(2J+1)(1+\delta_{j_{1}j_{2}})}}\times\Big<j_{1}j_{2}\text{\space}\frac{1}{2}\left(\frac{\text{-}1}{2}\right)\Big|J0\Big>
\end{equation}
 Note that the expression is separable and so, as indicated by Talmi
\cite{Talmi} it is easy to obtain the lowest state wave functions
(see Eq. 12.49). 
\begin{equation}
\psi^{J}=N\sum f(j_{1},j_{2})[j_{1}j_{2}]^{J}
\end{equation}

\section*{Wave Functions and $g$-Factors}
We had previously considered a 2-proton hole model fot $\mathrm{^{86}Kr}$
with the relevant orbits being $f_{5/2}$ and $p_{3/2}$ \cite{Zamick}, stimulated by experiments of Kubartzki et al.\cite{Kubartzki}.
We found that with the above SDI interaction and degenerate single
particle energies that the $g$ factors of the $2^{+}$ and $4^{+}$ states
were both equal to $g_{l}$ ,the orbital $g$ factor of a proton, The
free value of $g_{l}$ is one for the proton and zero for the neutron.
The spin part does not contribute. We will here consider neutrons
in the $g_{7/2}$ and $d_{5/2}$ shells and study more intensely the
effects of single particle splitting. (Note that a stray factor of
$2$ under the square root sign of $f(j_{1},j_{2})$ which appeared
in ref \cite{Zamick} has been removed here. We are now in accord with the
expression of Talmi \cite{Talmi}. The numerical results for $\mathrm{^{86}Kr}$ in
ref \cite{Zamick} are consistent with the definition of $f(j_{1},j_{2})$
in Eq{[}2{]} of this work).

This behaviour (vanishing $g_{s}$ contributions) is explained by the pseudo-spin symmetry ideas, developed by Hecht et al.\cite{Hecht} and Arima et al. \cite{Arima}. Insights into the origin of pseudo-spin symmetry via the Dirac equation has been giving by Ginocchio and Leviatan.\cite{Ginocchio}. Also relevant is the quasi-spin formulation of Kerman \cite{Kerman}
and Talmi's generalized seniority scheme \cite{Talmi}. A. Bohr et
al. showed how the spin operator transforms under a pseudo-spin transformation\cite{Bohr}.
From the pseudo-spin formulation one can map the orbits $g_{7/2}$ and $d_{5/2}$ into $f_{7/2}$ and $f_{5/2}$, thus having all the neutrons in one shell. We further note that the surface delta interaction is a quasi-spin conserving interaction. Note that the expression in Eq(2) depends on $j$ but not on $l$.

We consider $2$ neutron particles or $2$ neutron holes in either the $g_{7/2}$ or $d_{5/2}$ orbital. We choose $C_{0}$ to be to be $-0.2Mev$ We study how the $G(J)$ for $2$ holes or $2$ particles depends on the single particle splitting between the $g_{7/2}$ and $d_{5/2}$ orbits. We find that the wave function of the $2$ particle ($2$ holes ) state are :
\begin{equation}
BJ77[g_{7/2}g_{7/2}]^{J}+BJ75[g_{7/2}d_{5/2}]^{J}+BJ55[d_{5/2}d_{5/2}]^{J}
\end{equation}

We show the $g$ factors $G(J)$ for $2$ particle and also of $2$ holes in Tables
1 and 2. the first case we use free single particle $g$ factors $g_{l}=0$, $g_{s} =-3.826$ whilst in Table 2 , which we call quenched,
we have $g_{l} = -0.1$, $g_{s} =0.7$ $g_{s}(free) = -2.678$.

\begin{center}

\captionof{table}{$G(J)$ as a function of $E$ with bare input ($g_l=0$, $g_s=-3.826$)}
\begin{tabular}{|c|r r r|}
\hline 

\hline 
$E(MeV)$ & $G(2^{+})$ & $G(4^{+})$ & $G(6^{+})$\tabularnewline
\hline 
$-0.4$ & $0.386$ & $0.390$ & $0.323$\\
$-0.3$ & $0.362$ & $0.366$ & $0.247$\\
$-0.2$ & $0.314$ & $0.315$ & $0.142$\\
$-0.1$ & $0.210$ & $0.205$ & $0.052$\\
$0$ & $0$ & $0$ & $0$ \\

$+0.1$ & $-0.282$ & $-0.294$ & $-0.027$\\
$+0.2$ & $-0.496$ & $-0.542$ & $-0.042$\\
$+0.3$ & $-0.611$ & $-0.659$ & $-0.051$\\
$+0.4$ & $-0.669$ & $-0.708$ & $-0.056$\\
\hline 
\end{tabular}

\end{center}

\qquad

Here the single particle splitting is $E=\epsilon_{g_{7/2}}-\epsilon_{d_{5/2}}$. 
We verify that for $E=0$ we get $G(J) = g_{l}=0$ for the case of free neutron values. 
The spin part does not contribute.
Notice how rapidly the $G(J)'s$ change as a function of $E$. This super-sensitivity
makes it difficult to pind down the optimum values of the $G(J)'s$.
An added complication is that we should perhaps use renormalized values
of $g_{l}$ and $g_{s}$ in our analysis. This is shown in Table 2.
\begin{center}

\captionof{table}{$G(J)$ as a function of $E$ with quenched input ($g_l=-0.1$, $g_s=-2.678$)}
\begin{tabular}{|c|r r r|}
\hline 
$E(MeV)$ & $G(2^{+})$ & $G(4^{+})$ & $G(6^{+})$\tabularnewline
\hline

$-0.4$ & $0.160$ & $0.162$ & $0.118$\tabularnewline

$-0.3$ & $0.144$ & $0.146$ & $0.066$\tabularnewline

$-0.2$ & $0.112$ & $0.112$ & $0.005$\tabularnewline
 
$-0.1$ & $0.041$ & $0.038$ & $-0.065$\tabularnewline
 
0 & $-0.1$ & $-0.1$ & $-0.1$\tabularnewline

$+0.1$ & $-0.290$ & $-0.298$ & $-0.118$\tabularnewline

$+0.2$ & $-0.434$ & $-0.465$ & $-0.128$\tabularnewline
 
$+0.3$ & $-0.512$ & $-0.544$ & $-0.134$\tabularnewline

$+0.4$ & $-0.551$ & $-0.577$ & $-0.138$\tabularnewline
\hline 
\end{tabular}
\end{center}

Note that now for $E=0$ we get $G(J) = g_{l} = -0.1$. Again the renormalized spin part does not contribute.

If we take $E$ to be minus-infinity we get as asymptotic values for
2 holes : $G(2^{+})=G(4^{+}) = G(6^{+}) = g_{7/2}$.
This equals 0.4251 in the free case and 0.2533 in the quenched case.If
we take E to be plus infinity we also get all three $G$ factors to be
the same for 2 particles: $G(J)=g_{d_{5/2}}$. For a neutron this
equals $-0.7652$ in the free case and $-0.6156$ in the quenched case. 

In the nucleus $\mathrm{^{113}Sn}$ the $J= 7/2^{+} -J=5/2^{+}$ splitting
is $0.33 MeV$. If we identify this as a single hole nucleus and indeed
take the value of $E$ to be $-0.33$ then we obtain in the free case $G(2^{+})= 0.371$ , $G(4^{+}) =0.375$ and $G(6^{+})=0.274$. If we use the
renormalized values we get $G(2^{+}) =0.150$, $G(4^{+}) =0.153$
and $G(6^{+}) = 0.0845$. Notice that in all the cases that we have
considered, with E both positive and negative, we find that $G(2^{+})$
and $G(4^{+})$ are nearly equal. 

\section*{Linear Approximation}

So far we have considered 2 particles and 2 holes(12 particles) in
the model space $g_{7/2} d_{5/2}$ of neutrons. We now make the
speculation that we can obtain the $G(J)$ values for other $n$ by a linear
approximation. That is to say we assume $G=G(n=2)+ (G(n=12)-G(n=2) )/10{*}(n-2)$. This seems reasonable since the more $g_{7/2}$ neutrons we
add the more positive the $G(J)'s$ should become.

Note that if we go higher in neutron number, we get 2 new orbits for
 which we can use pseudo $LS$ coupling $s_{1/2}$ and $d_{3/2}$.
 If we play the same game, we can now include $A=114$ and $116$. With
 single particle energies degenerate, we get zero $g$ factors. In ref
 \cite{Allmond} the values for this $g$ factor are $+0.138(63)$, $0.0000(64)$. For completeness, we note that their values for $118$, $120$, $122$, $124$ are $0.000(77)$, $-0.0999(30)$, $0.000(48)$, and $-0.097(11)$.

Table 3, $G (J)$ as a function of number of valence particles in the
linear approximation for $E=0.3$ free values.

\qquad

\begin{minipage}[c]{0.45\linewidth}
\begin{center}

\captionsetup{width=0.8\linewidth}
\captionof{table}{Linear Approximation: $G(J)$ versus $n$ with $E=+0.3$ for 2 particles and $-0.3$ for 2 holes(bare input)}

\begin{tabular}{|c|r r r|}
\hline
$n$ & $G(2^{+})$ & $G(4^{+})$ & $G(6^{+})$\tabularnewline

\hline 
$2$ & $-0.611$ & $-0.659$ & $-0.051$\tabularnewline

$4$ & $-0.416$ & $-0.454$ & $0.009$\tabularnewline

$6$ & $-0.222$ & $-0.249$ & $0.068$\tabularnewline

$8$ & $-0.027$ & $-0.044$ & $0.128$\tabularnewline
 
$10$ & $0.167$ & $0.161$ & $0.187$\tabularnewline

$12$ & $0.362$ & $0.366$ & $0.247$\tabularnewline
\hline
\end{tabular}
\end{center}
\end{minipage}
\hfill
\begin{minipage}[c]{0.45\linewidth}
\begin{center}

\captionsetup{width=0.8\linewidth}
\captionof{table}{Linear Approximation: $G(J)$ versus $n$ with $E=+0.3$ for 2 particles and $-0.3$ for 2 holes(quenched input)}

\begin{tabular}{|c| r r r|}
\hline
$n$ & $G(2^{+})$ & $G(4^{+})$ & $G(6^{+})$\tabularnewline

\hline 
$2$ & $-0.512$ & $-0.544$ & $-0.134$\tabularnewline

$4$ & $-0.381$ & $-0.406$ & $-0.094$\tabularnewline

$6$ & $-0.250$ & $-0.268$ & $-0.054$\tabularnewline

$8$ & $-0.118$ & $-0.130$ & $-0.014$\tabularnewline
 
$10$ & $0.013$ & $0.008$ & $0.026$\tabularnewline

$12$ & $0.144$ & $0.146$ & $0.066$\tabularnewline

\hline

\end{tabular}
\end{center}

\end{minipage}

\section*{Column Vector}
As a supplement for Table 1 and Table 2, we list the eigenvalue and the eigenvector for each corresponding $E$ and $G$. Notice that for each  $E$ and $G$, there are 2 or 3 (the number depends on the dimensions of SDI-matrix) distinct eigenvalues and corresponding eigenvectors. We will pick the the eigenvector with smallest eigenvalue, for those are the correct ones. We use the same SDI-matrix but different $g_l$ and $g_s$ for Table 1 and 2. Therefore, two cases have the same eigensystem. 

\begin{center}

\captionof{table}{Eigenvalues and Eigenfunctions (column vectors) used to calculate $G(J)$}

\begin{tabular}{|c| r r r r r r r r r|}
\hline
\multicolumn{10}{|c|}{Eigensystem for $G(I=2)$}\\
\hline
$E$ &$-0.4$ & $-0.3$ & $-0.2$ & $-0.1$ & $0$&$0.1$&$0.2$&$0.3$&$0.4$\tabularnewline
$Eigenvalue$ &$-1.039$ & $-0.852$ & $-0.672$ &$-0.508$ & $-0.373$&$-0.282$&$-0.231$&$-0.203$&$-0.188$\tabularnewline
\hline 
$ g_{7/2} g_{7/2}$ & $0.966$ & $0.948$ & $0.914$ &$0.845$ & $0.714$&$0.536$&$0.381$&$0.280$&$0.215$\tabularnewline

$g_{7/2} d_{5/2}$ & $-0.177$ & $-0.212$ & $-0.258$ & $-0.312$ & $-0.350$&$-0.331$&$-0.274$ &$-0.218$&$-0.177$\tabularnewline

$d_{5/2} d_{5/2}$ & $0.189$ & $0.238$ & $0.314$ & $0.435$ & $0.606$&$0.777$&$0.883$& $0.935$&$0.960$\tabularnewline

\hline

\end{tabular}

\begin{tabular}{|c| r r r r r r r r r|}
\hline

\multicolumn{10}{|c|}{Eigensystem for $G(I=4)$}\\
\hline
$E$ &$ -0.4$&$ -0.3$ &$-0.2$ & $-0.1$ & $0$ & $0.1$ & $0.2$ & $0.3$&$0.4$\tabularnewline
$Eigenvalue$&$ -0.926$&$-0.737 $&$-0.555$ & $-0.389$ & $-0.255$ &$-0.164$ & $-0.116$&$ -0.095$&$-0.084$\tabularnewline
\hline 
$ g_{7/2} g_{7/2}$&$ 0.964$&$0.941$ & $0.891$ & $0.786$ & $0.606$ &$0.404$ & $0.250$&$0.164 $&$0.118$\tabularnewline

$g_{7/2} d_{5/2}$& $-0.244 $&$-0.310$ & $-0.410$ & $-0.542$ & $-0.639$ & $-0.430$ & $-0.350$&$-0.304$&$-0.227 $\tabularnewline

$d_{5/2} d_{5/2}$&$0.103$&$0.136$ & $0.194$ & $0.299$ & $0.474$ & $0.701$ & $0.867$&$0.938$&$0.967$\tabularnewline

\hline

\end{tabular}

\begin{tabular}{|c| r r r r r r r r r|}
\hline
\multicolumn{10}{|c|}{Eigensystem for $G(I=6)$}\\
\hline
$E$ &$-0.4$ & $-0.3$ & $-0.2$ & $-0.1$ & $0$&$0.1$&$0.2$&$0.3$&$0.4$\tabularnewline
$Eigenvalue$ &$-0.905$ & $-0.732$ & $-0.579$ & $-0.445$&$-0.326$&$-0.215$&$-0.108$ &$-0.003$&$0.100$\tabularnewline
\hline 
$ g_{7/2} g_{7/2}$ & $-0.892$ & $-0.800$ & $-0.654$ &$-0.498$ & $-0.378$&$-0.296$&$-0.240$&$-0.201$&$-0.172$\tabularnewline

$g_{7/2} d_{5/2}$ & $0.453$ & $0.600$ & $0.756$ & $0.867$ & $0.926$&$0.955$&$0.971$ &$0.980$&$0.985$\tabularnewline

\hline

\end{tabular}

 \end{center}

\section*{Additional Comments}

There are 2 known $g$ factors of relevance for odd-even nuclei. In $\mathrm{^{113}Sn}$
the $g$ factor for the $I=7/2^{+}$ state is $0.1737$ while in $\mathrm{^{109}Sn}$
the $g$ factor of the $I=5/2^{-}$ state is $-0.4316$. If we roughly identify
these as effective $g$ factors for $g_{7/2}$ and $d_{5/2}$ respectively
then for $E=0$ above we get an effective neutron $g_{l}$ equal to $-0.0425$.

The value $G(6^{+})$ in $\mathrm{^{110}Sn}$ as given by D.A. Volkov et al.
is $0.012$.\cite{Volkov} Most recently I.M. Allmond et al.\cite{Allmond} measured
many $g$ factors in the tin isotopes. The value of most interest in
this work is for $G(2^{+})$ in $\mathrm{^{112}Sn}$, namely $0.150(43)$. In our
work this would correspond to $n=12$ (or 2 holes). The Allmond et al.value \cite{Allmond} is close to what one gets for $I=7/2^{+}$ in $\mathrm{^{113}Sn}$
(in the single $j$ shell all $g$ factors are the same). It is also close
to what we get in our simple model here with renormalized values of
$g_{l}$ and $g_{s}$. This is for the choice $C_{0}=-0.2$ and $E=+0.3Mev$.

Note that if we go higher in neutron number, we get $2$ new orbits for which we can use pseudo $LS$ coupling  $s_{1/2}$ and $d_{3/2}$. If we play the same game, we can now include $A= 116$. With single particle energies degenerate, we get zero $g$ factors. In ref \cite{Allmond} the values for these $g$ factors is $0.0000(64)$. For completeness we note that their values for $A=114$, $118$, $120$, $122$, and $124$  are $+0.138(63)$, $+0.000(77)$, $-0.0999(30)$, $0.000(48)$, and $-0.097(11)$.

But just fitting one $g$ factor is not enough. A much more stringent
test would be to see if we can get $G(4^{+})$ and $G(6^{+})$ as
well. For example, in our surface delta model, we seem to get $G(4^{+})$
almost the same as $G(2^{+})$. This is either correct or incorrect,
and only experiment will tell us. Also in our work, we assume
that the $g$ factors vary linearly with $n$. Hence we have e.g. that $G(J)$
in $\mathrm{^{110}Sn}$ is smaller than $G(J)$ in $\mathrm{^{112}Sn}$.Only more experimentation
will tell us if this is correct or wrong. 

In a comprehensive paper by H. Jiang et al. \cite{Jiang}  one sees the opposite trend as one goes to lighter tin isotopes. namely an increase  in the $g$ factors as one approaches $\mathrm{^{102}Sn}$.   They they use an effective value of $g_l$ that is positive $g_l=+0.09$. This is strange, because all theories give a negative value  more or less equal and opposite to what they use. The effect of a positive $g_l$ is to make all the $g$ factors more positive.

It should also be pointed out that $E(7/2^{+}) -E(5/2^{+})$ varies
very strongly with mass number. The values for $A=113$, $111$, $109$, $107$, $105$, $103$, $101$ are respectively $-332.45$, $-154.48$, $+13.48$, $+151.2$, $+199.73$, $+168.0$ and $-172.2$ or$+172.3$ keV . The sign for A=101 is in dispute. This makes it difficult to know exactly which $E$ to choose. Note also the the relevant parameter is really $E/C_{0}$.

We feel that the main virtue of this simple model is that it provides
a basis for comparison with more sophisticated calculations and that
it provides insight into why many of the $g$ factors in this region
are so small. In this model when degenerate single particles are used
all $g$ factors of the even even nuclei for all spins vanish in the
range from $\mathrm{^{102}Sn}$ and $\mathrm{^{112}Sn}$ . Going away from zero the
$g$ factors in this model are super-sensitive to the parameters that
are used. This will also be true , although less transparent , in
more realistic calculations. So we see that in this region of the
periodic table both theorists and experimentalists are fighting zero.

\section*{A Brief Look at $\mathrm{^{126}Sn}$}
As a counterpoint to the case where we have pseudo $LS$ partners we
here consider a situation where this is not the case. We consider the
nucleus $\mathrm{^{126}Sn}$ where measurements were made by Kumbartzki et al.\cite{Kumbartzki}
and where one of us (LZ) was involved. Their result for $\mathrm{^{126}Sn}$
is $G(2+)=-0.25 (21)$. They also report the result for $\mathrm{^{124}Sn}$ as$ -0.36(17)$.
Th error bars are quite large. The relevant orbitals are $h_{11/2}$
and $d_{3/2}$. Empirically the 2 orbits are very nearly degenerate
e.g. in $\mathrm{^{125}Sn}$, $I={3/2}^{+}$ is $27.5 keV$ above $I=11/2^{-}$ and
in $\mathrm{^{127}Sn}$ the value is only $5.07 keV$. We we will therefore only
show the result for $E=0$.

We find $G(2) =-0.1096$,  $G(4)=-0.5812$, and $G(6)= -0.2758$. Note that we get
non-zero values for $E=0$ in contrast to what happens when one has pseudo
$LS$ pairs. Note also that $G(4)$ and $G(6)$ have not been measured and
we predict they should be substantially larger than $G(2)$.

The relevant wave functions are:

$I=0^{+}$ \quad $0.9737 h_{11/2} h_{11/2} + 0.2278 d_{3/2} d_{3/2}$

$I=2^{+}$ \quad $0.9865 h_{11/2} h_{11/2} + 0.4627 d_{3/2} d_{3/2}$

$I=4^{+}$ \quad $0.5487 h_{11/2} h_{11/2} + 0.8360 h_{11/2} d_{3/2}$

$I=6^{+}$ \quad $0.6757 h_{11/2} h_{11/2} + 0.7371 h_{11/2} d_{3/2}$

We see why $G(4)$ is much larger than $G(2)$. For $I=4$ one cannot have
a $d_{3/2}$ $d_{3/2}$ component with its accompanying positive $g$
factor. 

Very recently, the $E(J=7/2^+)-E(J=5/2^+)$  splitting in $\mathrm{^{107}Sn} $ has been measured by $g$. Cerizza  et al \cite{Cerizza}.  They get a value of $1551 keV$. This is in line with our choice of $E$.

\section*{Acknowledgement}
X.Y. acknowledges the support from the Rutgers Aresty program for
 2015-2016 academic year. We thank N. Koller, G. Kumbartzki, K-H. Speidel, S.J.Q. Robinson , Y.Y.Sharon  and  Jolie Cizewski for useful discussions.

\section*{Appendix }

We here give the $g$ factor used in our calculations. In Table 4 we
use the free values and in Table 5 the renormalized values , both
of which are given above. We use the formula

$G(J) = (g_{7/2}+ d_{5/2})/2 + (g_{7/2} - d_{5/2}) /(2
I(I+1)){*} ( 7/2{*}9/2-5/2{*}7/2)$.

Note that when $j_{1}=j_{2}$, $G(J)$ is independent of $J$. (see Table 6 and 7)

\qquad

\begin{minipage}[c]{0.4\linewidth}
\begin{center}

\captionsetup{width=0.8\linewidth}
\captionof{table}{Values of $G(J)$ for the basis states (bare input)}

\begin{tabular}{|c|r|}
\hline
Config. & $G(J)$\tabularnewline
\hline

$g_{7/2}g_{7/2}$ & $0.42511$\tabularnewline
 
$d_{5/2} d_{5/2}$ & $-0.76520$\tabularnewline

$g_{7/2}d_{5/2}$ $I=2$ & $0.52430$\tabularnewline

$g_{7/2}d_{5/2}$ $I=4$ & $0.03825$\tabularnewline

$g_{7/2}d_{5/2}$ $I=6$ & $-0.07085$\tabularnewline
\hline 
\end{tabular} 
\end{center}
\end{minipage}
\begin{minipage}[c]{0.4\linewidth}
\begin{center}

\captionsetup{width=0.8\linewidth}
\captionof{table}{Values of $G(J)$ for the basis states (quenched input)}

   \begin{tabular}{|c|r|}

\hline 
Config. & $G(J)$\tabularnewline
\hline
 
$g_{7/2}g_{7/2}$ & $0.16647$\tabularnewline
 
$d_{5/2}d_{5/2}$ & $-0.61564$\tabularnewline
 
$g_{7/2}d_{5/2}$ $I=2$ & $0.25331$\tabularnewline
 
$g_{7/2}d_{5/2}$ $I=2$ & $-0.07422$\tabularnewline
 
$g_{7/2}d_{5/2}$ $I=6$ & $-0.14774$\tabularnewline
\hline 
\end{tabular}

\end{center}
\end{minipage}

\qquad

\qquad

\qquad

\qquad

\qquad

\qquad

\end{document}